\documentstyle[epsf,prb,aps]{revtex} 

\begin{document}
\title{Sr$_3$CuIrO$_6$, a spin-chain compound with random 
ferromagnetic-antiferromagnetic interactions}

\author{Asad Niazi, E.V. Sampathkumaran,$^a$  P.L. Paulose,  
D. Eckert,$^*$  A. Handstein$^*$  and K.-H. M\"uller$^*$}

\address{Tata Institute of Fundamental Research, Homi Bhabha Road, 
Colaba, Mumbai - 400005, INDIA\\
$^*$Institut f\"ur Festk\"orper- and Werkstofforschung Dresden, Postfach
270016, D-01171 Dresden, GERMANY\\
$^a$E-mail address: sampath@tifr.res.in}

\maketitle

\begin{abstract}

Ac and dc magnetization and heat-capacity (C) measurements performed on the pseudo-one-dimensional compound Sr$_3$CuIrO$_6$ reveal a competition between antiferromagnetic (AF) and ferromagnetic (F) exchange couplings, as evidenced by frequency dependence of ac susceptibility and by the absence of a C anomaly at the magnetic transition. The value of the saturation moment (about 0.35 $\mu_B$/formula unit) is much smaller than expected for ferromagnetism from the two S=1/2 ions (Cu and Ir). Thus, this compound is not a ferromagnet in zero magnetic field, in contrast to earlier beliefs. Of particular importance is the finding that the value of the magnetic ordering temperature is sample dependent, sensitive to synthetic conditions resulting from deviations in oxygen/Cu content.  We propose that this compound serves as a unique model system to test theories on random AF-F interaction in a chain system, considering that this competition can be tuned without any chemical substitution.

\end{abstract}


\maketitle
\vskip 0.5cm

One-dimensional (1D) quantum spin systems have been of interest to physicists and chemists over many decades and this direction of research gained momentum in recent years. In particular, there are considerable theoretical efforts in understanding the thermodynamics of random ferromagnetic (F) - antiferromagnetic (AF) chain behavior.\cite{1,2,3} From the experimental point of view, to test such theoretical ideas, one brings out such a competition by introducing disorder effects {\it by chemical substitution} in pseudo-1D systems, for instance, the solid solution Sr$_3$CuIr$_{1-x}$Pt$_x$O$_6$, an often-quoted example (Refs. 1-4)  assuming that the Cu end-member is ferromagnetic below 19 K.\cite{4,5,6,7} However, the identification of a compound in which such disorder effects can be tuned {\it without} chemical substitution is desirable to test any such theory to avoid possible complications due to changes in the electronic structure.  In this article, we propose that the compound, Sr$_3$CuIrO$_6$, crystallizing in the monoclinically distorted form of K$_4$CdCl$_6$ structure (space group R$\bar3$c), characterized by the existence of magnetic chains of Cu-Ir-Cu separated by Sr ions serves itself as an example in this respect. At this juncture, we would like to state that this work is a continuation of our earlier efforts on isostructural Ru-based compounds.\cite{8,9}

We have investigated three different samples of the compound Sr$_3$CuIrO$_6$ systematically, with a reduction of Cu stoichiometry by 5 atomic percent for the third specimen (that is, Sr$_3$Cu$_{0.95}$IrO$_6$). All these specimens were prepared by a conventional solid state reaction method from the stoichiometric amounts of high purity ($>$ 99.9\%) SrCO$_3$, Ir metal, CuO and ZnO, intimately mixed under acetone. The specimen 1 was prepared by preheating at 800 $^\circ$C for 33 hours; the preheated powders were thoroughly ground again, pelletized and subjected to a further heat treatment at 1150 $^\circ$C for 72 hours with three intermediate grindings. The specimen 2 was also prepared with the same procedure except that the heat treatment at 1150 $^\circ$C was for 100 hours with three intermittent
grindings. The specimen 3 (Cu-deficient) was prepared by the same synthetic conditions as the specimen 1. The samples were characterized to be single phase (with monoclinic distortion) by x-ray diffraction (Cu K$_{\alpha}$).  We have measured dc magnetic susctibility ($\chi$) as a function of T (2-300 K) at different magnetic fields (H= 10, 100 and 5000 Oe) for the zero-field-cooled (ZFC) and field-cooled (FC) states of the specimens and ZFC isothermal magnetization (M) at 2 K by a commercial superconducting quantum interference device (Quantum Design) as well as by a vibrating sample magnetometer (Oxford Instruments); in addition, ac $\chi$ measurements were performed in the vicinity of magnetic transition temperature (T$_o$) at various frequencies.  The heat-capacity (C) measurements (2-40 K) were performed by a semi-adiabatic heat-pulse method; for sample 2, the data in the presence of external H were also taken by relaxation method.

We present the results of our investigations in Figs. 1-5. In sample 1, in the plot of $\chi$ versus T, measured at low fields, there is a steep rise below T$_o$ = 19 K, signalling the onset of magnetic ordering consistent with the earlier reports.  There is a corresponding sharp increase in the ac $\chi$ data as well; though we observe this both in real and imaginary parts, we show only the real part in Fig. 2 for the sake of clarity. A careful look at the available data reveals that the nature of the magnetic ordering is not as simple as the one reported (ferromagnetic?) in the literature.\cite{4,5,6} The plot of inverse $\chi$ versus T (Fig. 3) in the paramagnetic state at low temperatures (say, below 50 K)  of course extrapolates to a positive value of paramagnetic Curie temperature ($\Theta_p$) close to that of T$_o$, as if the magnetic ordering is of a ferromagnetic-type. At 2 K, we observe negligible variation of M (Fig. 4) for initial applications of H, that is, till about
few hundreds of gauss. This behavior may be connected with ferromagnetism
or spin-freezing. In the case of ferromagnetism pinning of domain walls
causes a weak variation of magnetization. But the observation that the virgin curve after ZFC deviates from the center of the hysteresis loop (see insets of Fig. 4) favours spin-freezing.  There is a sharp increase for H$>$ 300 Oe, followed by a tendency for saturation at higher fields. (These features are overall the same even at 10 K, though initial constancy of M is restricted to still lower fields). These features are typical of magnetic compounds exhibiting a spin-flip transition, thereby establishing that the zero-field magnetic structure in any case cannot be of a F-type.  The C data (Fig. 5, {\it vide infra}) in fact reveals that the long range magnetic ordering is absent. Considering that there is a tendency for saturation of M at higher fields, we conclude that the application of H however drives the system towards ferromagnetic ordering; also, the magnetic behavior is hysteretic, as there is a remnance of M evidenced by hysteresis loops at 2 K (Fig. 4, inset). We observe that the negligible field-dependence of low-field M discussed above is absent in the reverse cycle of the hysteresis loop; this
essentially means that there is a delicate balance between the history of
the sample and the magnetic behavior. In order to derive information on
the nature of the transition near T$_o$, we have also measured ac $\chi$
at various frequencies in the T range of interest in zero field as well as
in the presence of 1~kOe (Fig. 2)  below 40~K. In zero-field, there is a
frequency dependence of ac $\chi$ including the decrease of the intensity
of the peak as seen in Fig. 2; since the peak is broad, this dependence is
more clearly visible by the movement of the left-hand-side of the peak;
however, the data taken in the presence of 1 kOe at various frequencies
overlap, without exhibiting any frequency dependence. These observations
strongly endorse the points made above that in zero-field the compound
behaves like a spin-glass, whereas at higher fields such frustration
effects vanish. There is also a corresponding anomaly in the ZFC-FC dc
$\chi$ data: at low fields (say, at 10 Oe), ZFC and FC $\chi$ tend to
deviate near T$_o$, a characteristic of spin-glasses, and this divergence
is absent at higher fields, say at 5 kOe - consistent with above
conclusions.

With respect to sample 2, the anomalies due to magnetic ordering in the
$\chi$ data appear below 13 K. It is to be stressed that the observed
T$_o$ is significantly lower than that reported in the literature and in
sample 1.  This clearly establishes that T$_o$ is sample dependent for
this compound and prolonging the heat-treatment well beyond 24 hours after
each grinding thus depresses T$_o$ - a fact not known earlier in the literature - and we have confirmed this observation by repeated preparation of samples. In order to understand further the origin of this sample dependence, we have carried out energy dispersive x-ray analysis. We notice that the atomic ratios of metallic components are very close to the stoichiometry, but O contents are different in these two specimens; while in specimen 1, O content per formula unit is close to the ratio 6, specimen 2 contains excess oxygen (close to 7). It is therefore clear that prolonged heat treatment results in excess oxygen in-take apparently resulting in a
depression of T$_o$. It is of interest to focus future studies on the site
of extra oxygen. With respect to the nature of the magnetic ordering, we
face the same situation as in sample 1, with the anomalies due to magnetic
frustration being even more prominent. That is, unlike in sample 1, the ac
$\chi$ peak temperature itself is found to be highly sensitive to
frequency, shifting it upwards by as much as 1.6 K - much larger than in
sample 1 - as the ac frequency is increased.  While the plot of inverse
$\chi$ versus T at low temperatures (below 30 K) extrapolates to a
positive value of $\Theta_p$ as if the compound undergoes ferromagnetic
ordering, the low-field feature (nearly constant value)  in the plot of M
versus H at 2 K is the same as in sample 1.  The frequency dependence of
ac $\chi$ vanishes by a small application of an external dc magnetic field
(say, 1 kOe); also the plot of M versus H and the hysteresis loop at 2 K
look similar to that of sample 1 with comparable values (0.35 $\mu_B$ per
formula unit) of the saturation moment (extraplolated to zero field).  We
therefore interpret that the spin-glass-like state turns to ferromagnetic
ordering by the presence of a small magnetic field.

In order to understand how the vacancies at the magnetic Cu$^{2+}$ site
along the chain influence the magnetic behavior, we have also subjected
the third specimen (Cu deficient) to careful investigations. From the data
shown in Figs. 1-5, the features are broadly the same as in sample 1,
however with the reduction of T$_o$ (to about 17 K) compared to that of
sample 1. This finding establishes that the value of T$_o$ is very
sensitive to vacancies as well.  Otherwise, the interpretation of the
experimental data remain the same as for other two specimens.

We now present the results of C measurements, which throw additional light
on the nature of magnetism. The main finding is that there is no prominent
anomaly in the C data (Fig. 5) in any of the three samples at T$_o$,
except of a weak feature in sample 1.  For a comparison, we also show in
the Fig. 5 the data for the isomorphous Zn compound undergoing
well-defined AF ordering at the same temperature;\cite{6,10} it is
distinctly clear that the C jump at T$_o$ for the Zn compound is about 7
J/mol K, whereas the corresponding value for the sample 1 of the Cu
compound is less than 1 J/mol K. This finding renders strong evidence for
randomness of magnetic interactions in zero-field. In order to probe the
nature of the high field magnetic state, we have also taken the C data at
various fields for one of the samples (sample 2);  interestingly, the
feature due to magnetic ordering does not become prominent as expected for
perfect ferromagnets and C undergoes only a marginal increase just above
T$_o$; there is even a decrease well below T$_o$, which may reveal complex
nature of the magnetism of this compound (see inset of Fig. 5). Thus, the
overall change of entropy associated with the transition must be small
indicating a low-moment state even at high fields (possibly due to
ferrimagnetic-like ordering from interchain interactions as in the case of
the compounds reported in Ref. 11).  In support of this conclusion, the
saturation moment extrapolated from high fields (at 2 K) to zero H is
rather low (about 0.35 to 0.4 $\mu_B$/formula unit) compared to the value
expected if one assumes that there is a ferromagnetic ordering due to
half-integer spin at both Cu and Ir sites and to the value of the
effective moment (about 2.6 $\mu_B$ per formula unit)  in the paramagnetic
state, say in the linear region 250-300 K (Fig. 3).

While all the data discussed above present evidence for the existence of a
competition between F and AF interactions in the magnetically ordered
state, we do not prefer to call this compound a conventional spin-glass as
the experimental features have been known in many situations where
magnetic frustrations are favored by disorder.\cite{12} The nature of the
magnetic interaction in the paramagnetic state, particularly whether this
competition persists above T$_o$, remains to be answered. The plot of
inverse $\chi$ versus T indeed offers an affirmative answer to this
question. A common feature to all these three specimens is that this plot
in the entire range of the paramagnetic state (Fig. 3) is found to be
highly non-linear. Since this compound is an insulator, such a
non-linearity cannot arise from (temperature independent) Pauli
paramagnetic contribution. According to Ref. 3, it has to be attributed to
a competition between AF and F exchange couplings.  From the linear region
of the plot of inverse $\chi$ versus T above 200 K, the sign of $\Theta_p$
is found to be negative with a large magnitude (about -60 K); this
establishes that the interaction between the spins are dominantly AF. 
With decreasing T, the spin correlation length
increases\cite{2} with increasing domination of ferromagnetic coupling as
indicated by the positive sign of $\Theta_p$ at low temperatures. Thus
there is a competition between AF and F exchange interactions varying with
temperature even in the paramagnetic state.  As evidenced by the existence
of magnetic ordering at low temperatures, there is a crossover in magnetic
dimensionality with decreasing T.

Summarizing, this work establishes that the compound, Sr$_3$CuIrO$_6$, is
not a ferromagnet in zero magnetic field in sharp contrast to earlier proposals, but exhibits spin-glass like characteristics, which however can be suppressed by the application of a magnetic field. We thus have identified a chain compound, viz., Sr$_3$CuIrO$_6$, in which there is a competition between AF and F exchange interactions among the magnetic ions in the chain above and below magnetic transition temperature. This compound also offers an opportunity to tune this competition just by varying synthetic conditions and hence a nice candidate for further studies for better understanding of the theoretical predictions of such a magnetic competition in low dimensional systems. The results reveal that this compound is an interesting magnetic material, rich in physics and sufficiently motivating for further studies.\cite{13,14}

Acknowledgements:  This work in Germany was supported by the Deutsche Forschungsgemeinschaft within the SFB463. One of us (EVS) would like thank C. Laubschat for an invitation to Dresden, where a part of ac susceptibility studies were carried out. We are also grateful to Dr. Sudhakar Reddy
(Germany) for elemental analysis of the samples.

\newpage

\begin{figure}[htbp]
\centerline{\epsfxsize=15cm{\epsffile{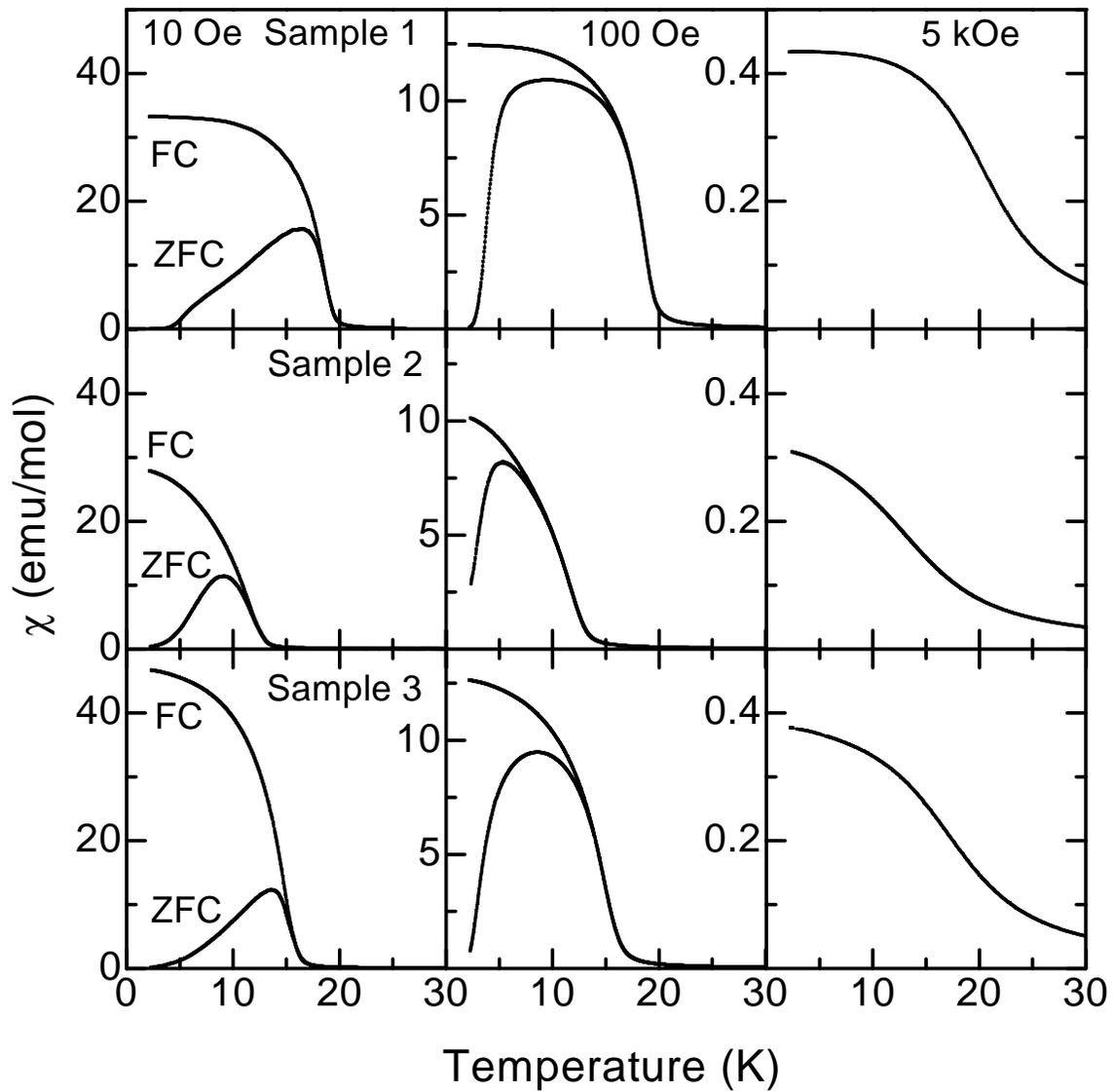}}}
\caption{Temperature dependence of magnetic susceptibility ($\chi$) below
30 K measured in the presence of different magnetic fields (10, 100 and
5000 Oe) for the zero-field-cooled (ZFC) and field-coled (FC) states of
the three specimens, 1, 2 and 3 of Sr$_3$CuIrO$_6$. The ZFC-FC curves for
H= 5000 Oe overlap.}
\end{figure}

\begin{figure}[htbp]
\centerline{\epsfxsize=15cm{\epsffile{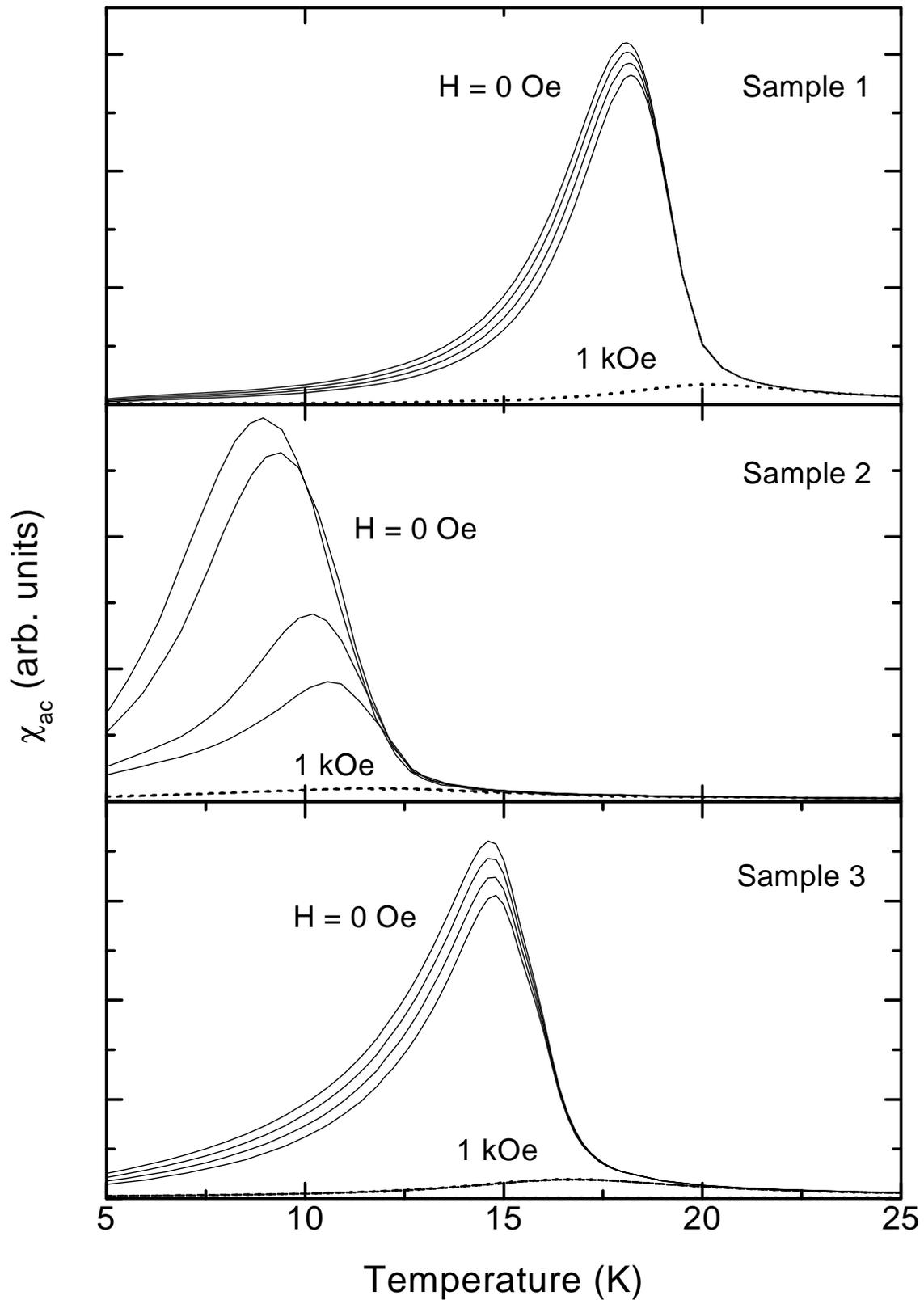}}}
\caption{Real part of ac susceptibility as a function of temperature for
the three specimens of Sr$_3$CuIrO$_6$ at different frequencies (for the
curves shown for each sample, top to bottom: 3, 10, 100, 1000 Hz) in the
region of magnetic ordering;. The data were also taken at all these
fequencies in the presence of a magnetic field of 1 kOe and the curves for
different frequencies are found to lie one over the other.}
\end{figure}

\begin{figure}[htbp]
\centerline{\epsfxsize=15cm{\epsffile{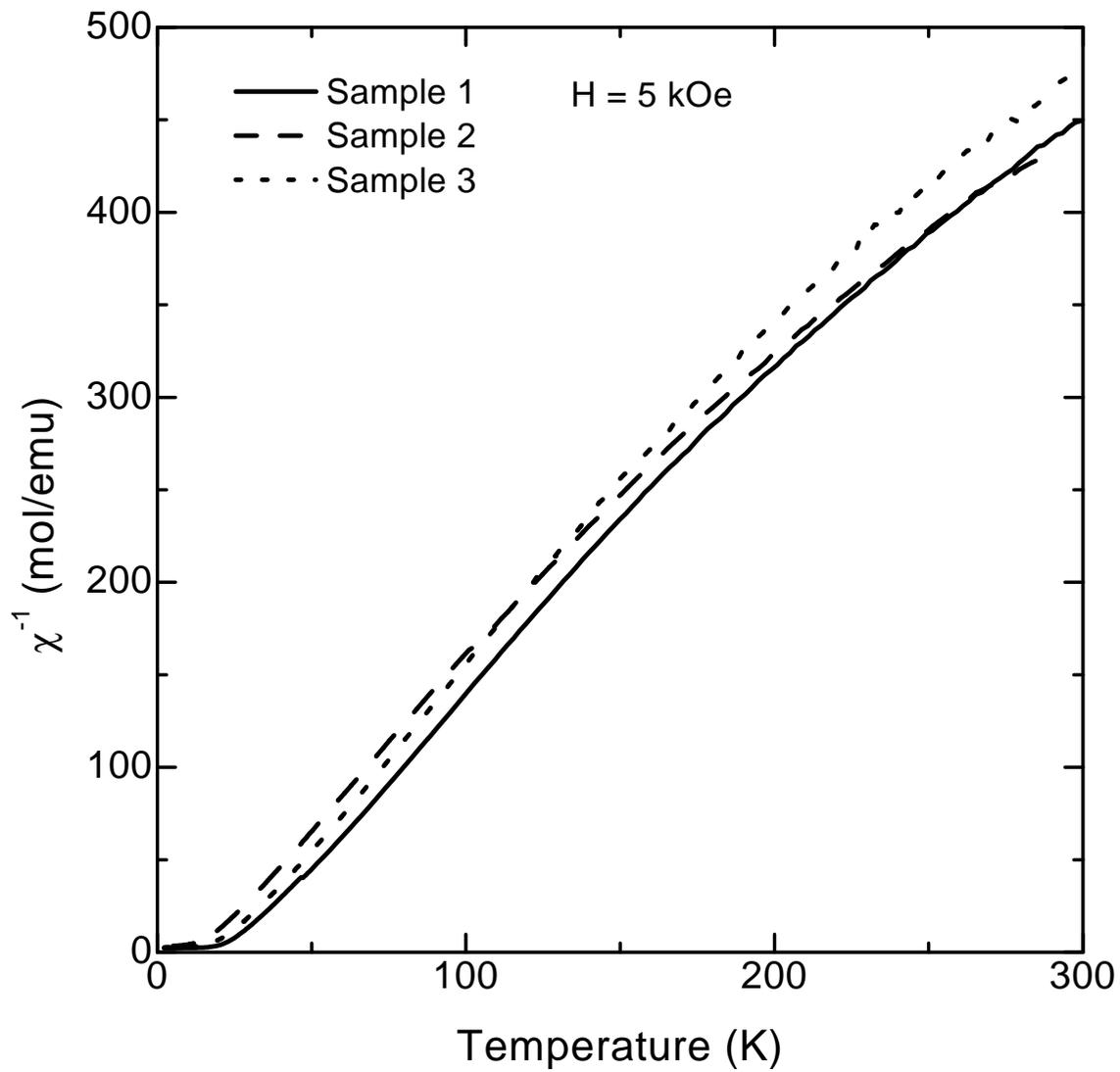}}}
\caption{Inverse susceptibility as a function of temperature measured for
all the 3 samples of Sr$_3$CuIrO$_6$ (zero-field-cooled state). }
\end{figure}

\begin{figure}[htbp]
\centerline{\epsfxsize=15cm{\epsffile{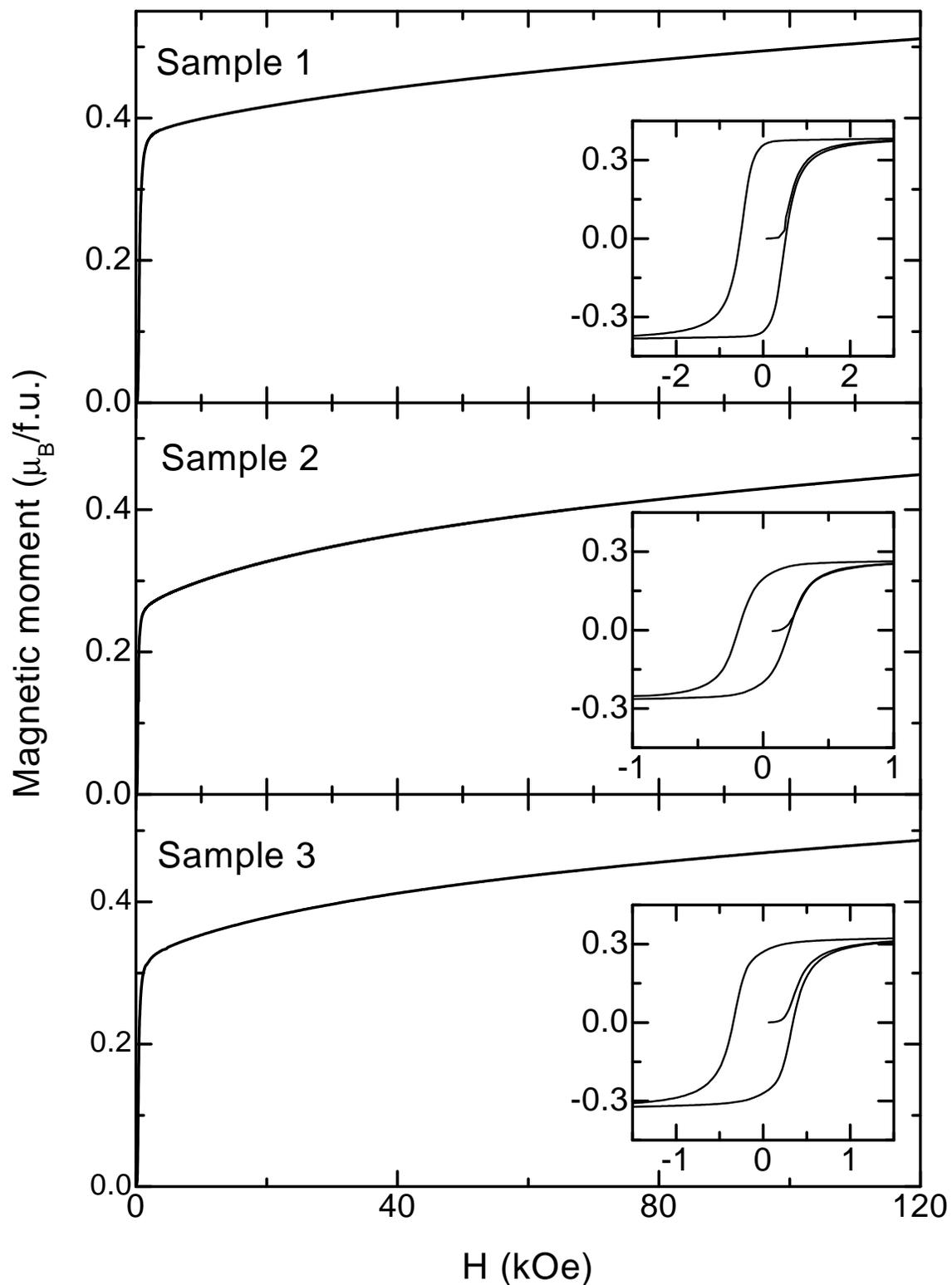}}}
\caption{Isothermal magnetization at 2 K (after ZFC) as a 
function of magnetic field
up to 120 kOe for the three specimens of Sr$_3$CuIrO$_6$. Low field
magnetic hysteresis loops are shown in the insets.}
\end{figure}

\begin{figure}[htbp]
\centerline{\epsfxsize=15cm{\epsffile{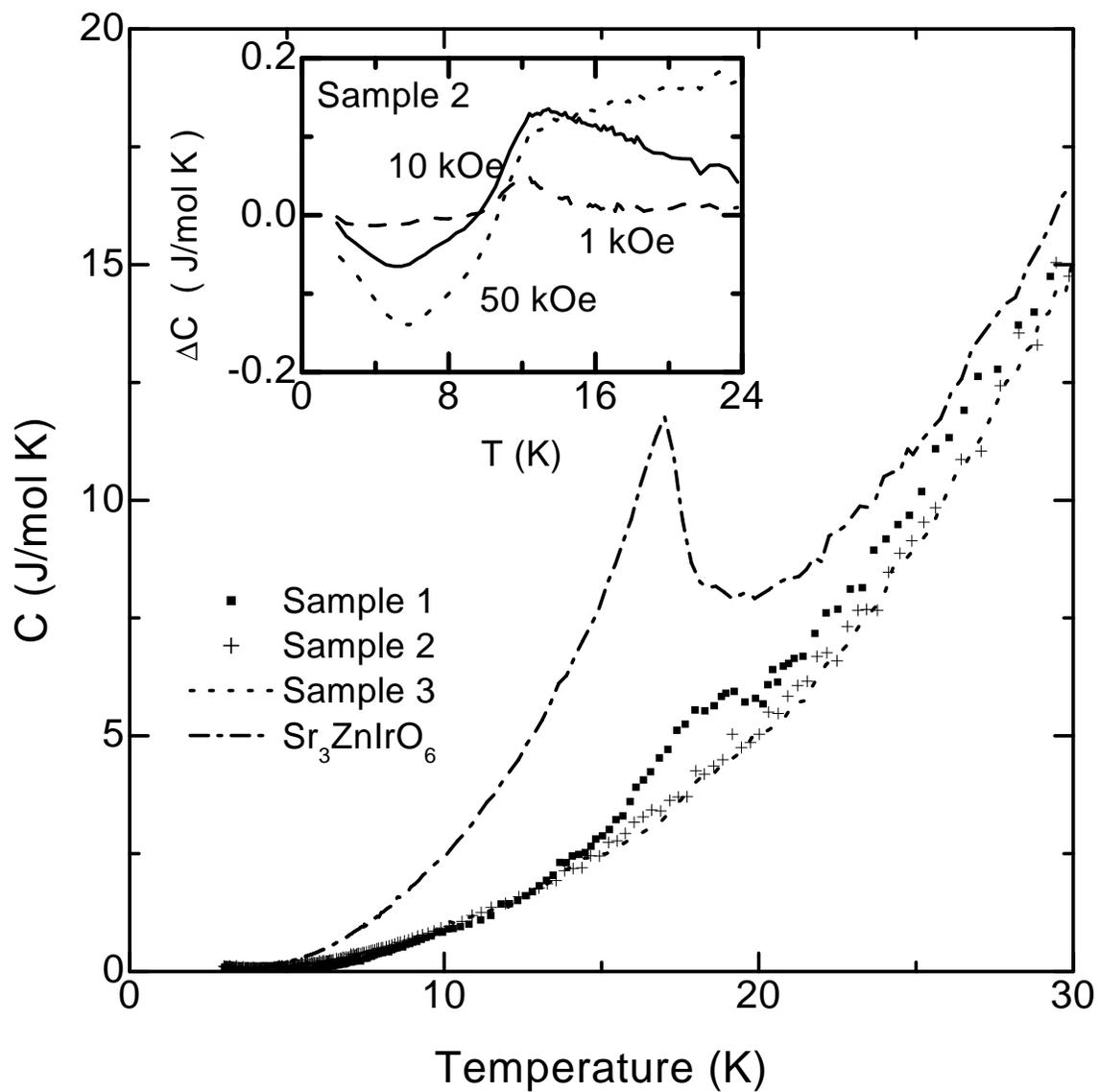}}}
\caption{Heat capacity as a function of temperature for all the three
specimens of Sr$_3$CuIrO$_6$. The data for the corresponding Zn compound
are also shown. For the specimen 2, the data were also taken in the
presence of externally applied magnetic fields and the change in C
($\Delta$ C= C(H)-C(0)) is shown in the inset.}
\end{figure}      

\end{document}